# An introduction to programming Physics-Informed Neural Network-based computational solid mechanics

Jinshuai Bai[a,b,d], Hyogu Jeong[a], C. P. Batuwatta-Gamage[a], Shusheng Xiao[a], Qingxia Wang[c,d], C.M. Rathnayaka[e], Laith Alzubaidi[a,b], Gui-Rong Liu[f], Yuantong Gu[a,b,*]

[a] School of Mechanical, Medical and Process Engineering, Queensland University of Technology, Brisbane, QLD 4000, Australia

[b] ARC Industrial Transformation Training Centre—Joint Biomechanics, Queensland University of Technology, Brisbane, QLD 4000, Australia

[c] Centre for Applied Climate Sciences, University of Southern Queensland, Toowoomba, QLD 4305, Australia

[d] School of Civil Engineering, The University of Queensland, Brisbane, QLD 4072, Australia

[e] School of Science, Technology and Engineering, University of the Sunshine Coast, Petrie, QLD 4502, Australia

[f] Department of Aerospace Engineering and Engineering Mechanics, University of Cincinnati, OH 45221, USA

* Corresponding authors.

*E-mail addresses:* yuantong.gu@qut.edu.au (Y. Gu),

## Abstract

Physics-informed neural network (PINN) has recently gained increasing interest in computational mechanics. This work extends the PINN to computational solid mechanics problems. Our focus will be on the investigation of various formulation and programming techniques, when governing equations of solid mechanics are implemented. Two prevailingly used physics-informed loss functions for PINN-based computational solid mechanics are implemented and examined. Numerical examples ranging from 1D to 3D solid problems are presented to show the performance of PINN-based computational solid mechanics. The programs are built via Python with TensorFlow library with step-by-step explanations and can be extended for more challenging applications. This work aims to help the researchers who are interested in the PINN-based solid mechanics solver to have a clear insight into this emerging area. The programs for all the numerical examples presented in this work are available at https://github.com/JinshuaiBai/PINN_Comp_Mech.

# 1. Introduction

Computational mechanics is the discipline that applies numerical techniques to solve mechanics problems. The majority of problems in mechanics can be represented using partial differential equations (PDEs) in different forms depending on the type of problem. Analytically solving these types of PDEs is found difficult for irregular problem domains. Thus, numerically solving PDEs becomes the central focus of computational mechanics. Numerous numerical approaches, such as the finite element method (FEM) and the element free Galerkin (EFG), or meshfree methods have been developed [1]. The most widely used method is the FEM not only in research but also in a wide range of industrial applications [2].

In recent years, deep learning (DL), i.e., multilayer perceptron that is fully connected artificial neural network (ANN), has attracted great attention in computational mechanics and provided alternative ways for solving mechanics problems. ANNs can be equipped with multiple hidden layers with neurons, providing them with powerful learning capability. As long as there is a sufficient amount of quality data or some form of information that can be used to train an ANN properly, it can become a powerful tool for a wide range of problems, including image processing [3, 4], biological predictions [5], and non-smooth dynamics [6]. The quality data is, however, often difficult to acquire.

A fully trained ANN can also provide a powerful tool to unveil the relationships between field variables of mechanics problems. Based on the training data, optimisers can iteratively improve the performance of ANNs by tuning the parameters. Many neural network-based computational mechanics has been proposed [7, 8]. However, in most mechanics problems, only a limited number of observed or measured data are available. For example, quantifying inside substance and structures are hard to obtain even with state-of-the-art equipment [9]. It is very challenging, if not impossible, to obtain a well-developed and labelled database for computational mechanics. ANNs trained with insufficient data are susceptible to severe failures, such as low prediction accuracy and poor generalisation [10]. Therefore, data scarcity greatly hinders the applications of DL in computational mechanics.

Recently, the physics-informed neural network (PINN) has been proposed by Raissi et al. [11], in which the physics laws and equations can be seamlessly integrated into training ANNs. In PINN, physics laws are transferred into training data. Compared to data-driven DL, PINN leverages physics laws as the remedy for insufficient data. In this manner, PINN can achieve better performance than data-driven DL when facing data scarcity. Up to now, a great number of studies have been conducted for PINN. Meng et al. [12], and Dong and Li [13] introduced domain decomposition techniques, which segment complex problems into subdomains, to PINN. With such techniques, PINNs are now more powerful in coping with extreme PDEs. Wang et al. [14, 15] studied the training pathology of PINN and migrated the neural tangent kernel (NTK) theory to PINN. Based on the numerical findings, an annealing learning rate and adaptive training algorithm were proposed for PINN so that PINNs can be more effective for problems with high-frequency features. Sukumar and Srivastava [16] proposed the distance function method that can exactly impose boundary and initial conditions on PINN. Yang et al. [17] integrated Bayesian distribution into PINN and proposed the B-PINN, which can evaluate the noisy data during training processes. Furthermore, Wang et al. combined the Fourier feature embeddings with PINN and tailored neural network structures. With the tailored neural networks, PINN can be used to deal with multiscale feature problems. Kharazmi et al. [18] proposed a variational type PINN, which trains neural networks to find the stationary points of functional loss functions. Moreover, several mixed-form loss functions have been proposed [19], aiming at fusing the benefits of various loss function types. It has been proven that PINN is effective for solving PDEs. For more recent developments of PINN, readers are referred to [20].

Physics-informed neural network-based computational mechanics has become one of the most popular topics in computational mechanics. Haghighat et al. [21] proposed a PINN-based framework for predicting field variables (such as displacement and stress) in elastostatic and elastoplasticity problems. In their study, ANN is applied to approximate the displacement field of mechanics problems. In addition, the collocation approach is used to incorporate governing equations and boundary conditions in the physics-informed loss function in order to assess the performance of ANN. The residuals of the ANN approximated displacement are minimised using optimisers until convergence. Samaniego et al. [22] leveraged PINN for computational mechanics from the energy point of view. In their work, the principle of minimum potential energy is applied as the physics laws for static problems. Optimisers are used to seek the stationary state of the total potential energy for static mechanics problems. Up to now, various PINN-based computational mechanics have been proposed for numerous applications, including fluid mechanics, vibrations, and fracture, to name but a few [9, 19, 23-33].

Compared to traditional computational mechanics methods, PINN-based computational mechanics has the following advantages [21, 22, 32, 34]:

- Physics-informed neural network provides a powerful tool for solving nonlinear systems. In this manner, PINN-based computational mechanics can easily deal with mechanics problems with nonlinearity, such as large deformation and material nonlinear problems. Those problems are considered to be challenging for traditional computational mechanics methods.
- Partial differential terms can be analytically obtained through automatic differentiation instead of spatial discretisation schemes and approximation methods in traditional computational mechanics. In this manner, PINN-based computational mechanics frameworks are less likely to be affected by the mesh quality and the distortion problems in mesh-based, or the arbitrary particle distribution and the boundary truncation issues in meshfree methods.
- Physics-informed neural network-based computational mechanics has great potential for solving inverse problems.
- Physics-informed neural network-based computational mechanics is easy to implement. Open libraries for coding PINN are available online and are easy to use.

This study presents the details of the basic conceptions and implementations of PINN-based computational solid mechanics. Comparisons between different types of physics-informed loss functions are summarised. It has been proven that PINNs with the collocation loss function can produce accurate predictions for both displacement and stress fields, for both the equilibrium equation and boundary conditions are enforced in the physics-informed loss function. PINNs with the energy-based loss function can produce good displacement predictions but have large errors for the stress predictions. This is because the energy-based loss function indirectly imposes the governing equation. However, the energy-based loss function requires lower partial differential terms than the collocation loss function. Hence, the energy-based loss function is easier to implement and computationally more efficient than the collocation loss function.

The aim of this work is to help researchers and engineers comprehend PINN-based computational mechanics and its programming. To this end, fundamental PINN-based computational solid mechanics programs are available with step-by-step explanations for 1D, 2D, and 3D problems. The programs are written in Python with the TensorFlow library [35], which is one of the most popular DL libraries. Besides, readers can easily extend the program to challenging solid mechanics problems, such as geometric nonlinearity [19, 36], hyperelastic [37], and fracture problems [38].

This paper is organised as follows. Section 2 introduces the essential conceptions of PINN-based computational solid mechanics. In Section 3, a 1D problem is conducted to elucidate the implementations of the PINN-based computational mechanics. Besides, the program corresponding to the 1D problem is also provided in detail. In Sections 4 and 5, the extensions of PINN-based computational solid mechanics for 2D and 3D problems are provided with discussions. Section 6 summarises and provides further perspectives. The program for all the numerical examples presented in this paper is available at https://github.com/JinshuaiBai/PINN_Comp_Mech.

## 2. PINN-Based Computational Solid Mechanics

In this section, the essential conceptions of PINN-based computational solid mechanics are introduced. First, governing equations and boundary conditions in solid mechanics are recalled. Next, a brief introduction of PINN and its application to solve solid mechanics problems are elucidated. Finally, a boundary condition imposition technique for the PINN-based computational solid mechanics is presented.

*2.1. Governing equations in solid mechanics*

In solid mechanics, the governing equation for linear elastic problems is as follow

$$\sigma_{\alpha\beta,\beta} + f_\alpha = 0, \quad x \in \Omega, \tag{1}$$

where $\sigma$ is the Cauchy stress tensor, $f$ is the body force per unit mass. Under the small deformation assumption, the stress tensor is calculated by

$$\sigma_{\alpha\beta} = \lambda \delta_{\alpha\beta} \varepsilon_{\gamma\gamma} + 2\mu \varepsilon_{\alpha\beta}, \tag{2}$$

$$\varepsilon_{\alpha\beta} = \frac{1}{2}(u_{\alpha,\beta} + u_{\beta,\alpha}), \tag{3}$$

where $\lambda$ and $\mu$ are the Lamé constants. $\delta_{\alpha\beta}$ is the Kronecker delta function. $\varepsilon$ is the strain tensor and $u$ is the displacement. The above governing equations are closed by boundary conditions, which can be written as

$$u_\alpha = \bar{u}_\alpha, \quad x \in \Gamma_u, \tag{4}$$

$$\sigma_{\alpha\beta} n_\beta = \bar{t}_\alpha, \quad x \in \Gamma_t, \tag{5}$$

where $\bar{u}$ and $\bar{t}$ denote the displacement and traction force on the corresponding boundaries, respectively, and *n* denotes the unit outward normal vector on the corresponding boundaries [39].

## 2.2. Physics-informed Neural Network for solid mechanics

PINN comprises two main components, i.e., ANN and physics-informed loss function. Herein, details of ANN and physics-informed loss functions are introduced, respectively.

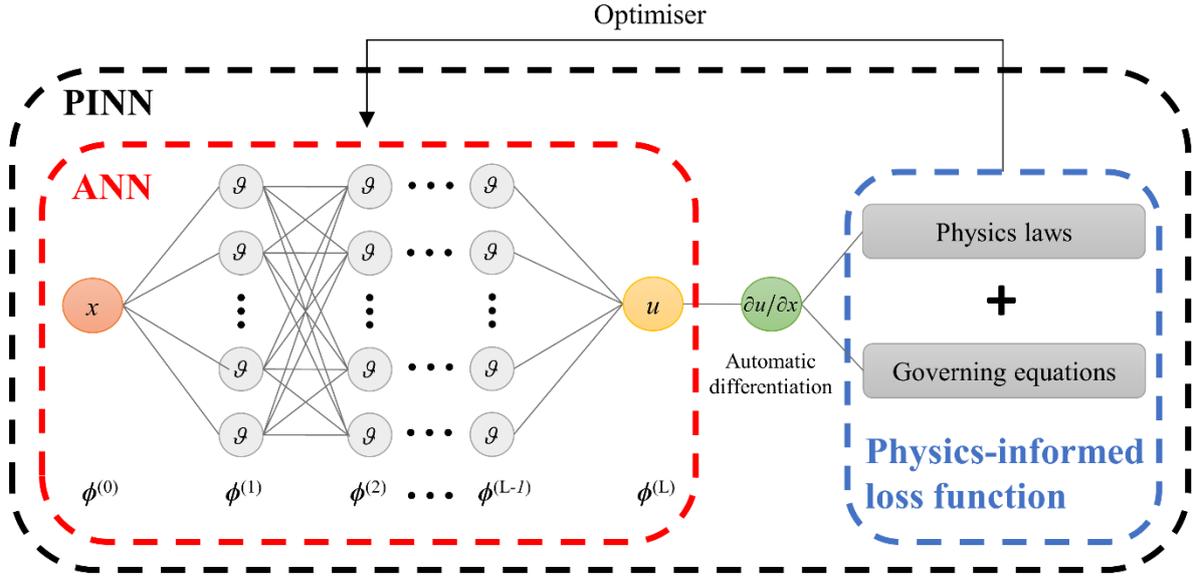

Fig. 1. A physics-informed neural network comprises two main components, i.e., artificial neural network (ANN) and physics-informed loss function. An example of an *L*-layer ANN is shown in the red dash box. *x* and *u* denote the input and output of the ANN, respectively. $\vartheta$ denotes the activation function used in the ANN. The physics-informed loss function is shown in the blue dash box. The physics-informed loss function is formulated by the physics laws and governing equations of the investigating systems. Note that partial differential terms, which are widely seen in physics laws and governing equations, are able analytically obtained via automatic differentiation.

### 2.2.1. Artificial Neural Network

The artificial neural network is a powerful bionic computing system that is inspired by the biological brain [40]. It consists of multiple neural network layers with artificial neurons, an example of an ANN is shown in Fig. 1. Specifically, the first layer is the input layer while the last layer is the output layer. In an ANN, the adjacent neural network layers are connected with each other via weights, biases and activation functions [41]. When using ANN, input information is fed into the neural network through the input layer, and then propagated through the adjacent layers. Finally, the prediction of the ANN is output through the output layer. An L-layer ANN can be mathematically expressed as [10]

$$\begin{aligned}
\phi^{(0)} &= x; \\
\tilde{\phi}^{(l)} &= w^{(l)}\phi^{(l)} + b^{(l)}; \\
\phi^{(l+1)} &= \vartheta\left(\tilde{\phi}^{(l)}\right); \\
u = \phi^{(L)} &= w^{(L-1)}\phi^{(L-1)} + b^{(L-1)},
\end{aligned} \tag{6}$$

where $w^{(l)}$ and $b^{(l)}$ are the $l$-th layer's weights and biases in the ANN, respectively. $\vartheta$ denotes the activation function, which provides nonlinear features to the neural network [41].

Referred to the *universal approximation theorem* [42], an ANN is capable of approximating any Borel measurable functions by changing the values of weights and biases [43]. Therefore, ANNs are widely used to study and capture the underlying relationship between inputs and outputs. In the PINN-based computational solid mechanic, ANN is applied to approximate the displacement field by using the coordinate information. For different solid mechanics problems, training algorithms are used to seek different sets of the optimal weights and biases inside the ANN [10].

2.2.2. Physics-informed loss function

In DL, the loss function quantifies the performance of ANNs with current weights and biases. Generally, it calculates the overall differences between the ANNs' output and the ground truth data (such as experimental data and observations). Based on current loss, training algorithms are applied to improve the performance of ANNs by modifying the weights and biases. In PINN, the physics-informed loss function is applied to estimate the performance of ANNs. As indicated by its name, the physics-informed loss function is formulated by physics laws, which can effectively govern how the variables change under the investigating systems [44]. By using such a loss function, the physics laws provide additional knowledge in training ANNs. In this manner, PINN can achieve favourable accuracy than ANN which is only trained by ground truth data.

Additionally, by applying specific activation functions such as tanh, sigmoid and mish [45], ANNs provide differentiable mathematical mappings from inputs to outputs. Thus, partial differential terms in the physics and governing equations can be analytically obtained through automatic differentiation [46], while traditional computational mechanics requires numerical schemes to approximate the partial differential terms. In traditional computational mechanics, the schemes for approximating partial differential terms rely on the surrounding information. When the surrounding information is inadequate, such as in cases of uneven particle distribution or near-boundary conditions, the accuracy of the approximation will be greatly impacted. These problems can be greatly mitigated through neural network mappings with automatic differentiation [32].

In PINN-based computational mechanics, there are mainly two kinds of physics-informed loss functions, i.e., the collocation loss function and the energy-based loss function. Herein, brief introductions for them are summarised as follows:

- **Collocation loss function**

    The collocation loss function is the most straightforward loss function in PINN-based computational solid mechanics. It sums up the Mean Square Error (MSE) from the governing equation and traction boundary condition at every sample point, which reads

    $$\mathcal{L} = \mathcal{L}_g + \mathcal{L}_t, \tag{7}$$

    where $\mathcal{L}_g$ and $\mathcal{L}_t$ respectively denote the loss terms from the governing equation and boundary condition. By applying the Eq. (1) and (5), the collocation loss function can be written as

    $$\mathcal{L} = \frac{1}{n}\sum_{i=1}^{n}\left(\sigma_{\alpha\beta,\beta}^{i} + f_{\alpha}^{i}\right)^{2} + \frac{1}{m_t}\sum_{i=1}^{m_t}\left(\sigma_{\alpha\beta}^{i} n_{\beta} - \bar{t}_{\alpha}^{i}\right)^{2}, \tag{8}$$

    where $n$ is the total number of sample points, and $m$ is the number of sample points on the traction boundary. The idea of the collocation loss function is to minimise the physics residuals at every sample point by modifying the weights and biases in the ANN. Despite the simplicity of implementing the collocation loss function, the usage of the collocation loss function may induce the bias training issue. Since the collocation loss function simply sums up residuals from different physics, the magnitude differences of the residuals can be significant. During the training process, the optimiser may pay more attention to minimising the relatively larger loss terms while neglecting the others. To address this problem, techniques including adaptive learning and dimensionless formulation can be applied. For more details, readers can refer to [15, 47].

- **Energy-based loss function**

    Based on the principle of minimum potential energy, the energy-based loss function is proposed based on the variational PINN [48]. In the energy-based loss function, the overall potential energy of the solid system is used as the loss function

    $$\mathcal{L} = E_{in} - E_{ex}, \tag{9}$$

where $E_{in}$ and $E_{ex}$ are the internal potential energy and potential energy of the external force, respectively. The two potential energies can be obtained through

$$E_{in} = \int_\Omega \frac{1}{2}\sigma_{\alpha\beta}\varepsilon_{\alpha\beta}\mathrm{d}V, \tag{10}$$

$$E_{ex} = \int_{\Gamma_t} u_\alpha \bar{t}_\alpha \mathrm{d}\Gamma_t. \tag{11}$$

Therefore, the energy-based loss function can be written as

$$\mathcal{L} = \int_\Omega \frac{1}{2}\sigma_{\alpha\beta}\varepsilon_{\alpha\beta}\mathrm{d}V - \int_{\Gamma_t} u_\alpha \bar{t}_\alpha \mathrm{d}\Gamma_t. \tag{12}$$

The idea of the energy-based loss function is to find a neural network mapping that reaches the stationary point of the overall potential energy. Compared to the collocation loss function, the energy-based loss function naturally unifies the unit of all loss terms into the energy unit. Therefore, the energy-based loss function greatly alleviates the bias training issue [26]. Additionally, as observed from Eq. (12), the energy-based loss function only requires the first-order derivative of the displacement $u$, while the collocation loss function requires the second-order derivative of the displacement $u$. Thus, implementing the energy-based loss function can have a simpler way of coding and is computationally more efficient than the collocation loss function [26]. Nevertheless, the energy-based loss function suffers inaccurate strain and stress fields, because it does not explicitly embed the equilibrium equation, which describes the stress balance condition [36].

Additionally, Henkes et al. [43], and Fuhg and Bouklas [19] tried to combine the above two loss functions. However, the collocation loss function and the energy-based loss functions have different physical units. In another word, these two loss terms are physically not able, to sum up. Thus, by using such mixed-form loss functions, PINNs perform unstably and can always generate ridiculous predictions. Artificially weights are required to carefully balance the different loss terms.

*2.3. Fixed boundary conditions imposition*

Imposing the fixed boundary condition is important in PINN-based computational solid mechanics. It is worth noting that, with boundary condition imposition techniques, fixed boundary conditions can be naturally satisfied. In this manner, the physics-informed loss function can be simplified, resulting in significantly decreasing the complexity of the training process and training expense. Currently, various boundary condition imposition techniques (such as distance function [16] and tailored ANN structure [22]) have been proposed for PINN. In this study, we adopt the boundary imposition technique introduced by Samaniego et al. [22] to deal with fixed boundary conditions. Considering a fixed boundary condition as follow

$$u = 0, (x = a). \tag{13}$$

To exactly implement the fixed boundary condition, we tailored the output layer of the ANN as

$$u = (x - a) \cdot \hat{u}, \tag{14}$$

where $\hat{u}$ is the direct output from the original ANN, and $u$ is the actual output after the tailored output layer. Consequently, $u$ satisfies the fixed boundary condition in Eq. (13).

## 3. PINN-based Computational Solid Mechanics in One Dimension

### 3.1. Stretching rod

Herein, a 1D stretching rod problem is considered to illustrate the implementation of the PINN-based computational solid mechanics. The configuration of the problem is given in Fig. 2. The length of the rod is 1 m and Young's modulus of the material is $E = 10$ Pa. A stretching force, $\bar{t} = 1$ N, is applied at the right end of the rod. In this manner, the equilibrium equation and the corresponding boundary conditions can be given by

$$\begin{aligned} \sigma_{x,x} &= 0, x \in [0,1] \\ u(x=0) &= 0, \\ \sigma_x(x=1) &= 1. \end{aligned} \tag{15}$$

From the energy point of view, the overall potential energy of the problem can be written as

$$\Pi = \int_0^1 \frac{1}{2} \sigma_x \varepsilon_x \mathrm{d}x - \bar{t} \cdot u(x=1). \tag{16}$$

Finally, the analytical solution to this problem is

$$u(x) = \frac{x}{E}. \tag{17}$$

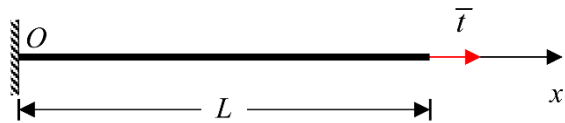

Fig. 2. Configuration of the stretching rod problem. The arrow in red denotes the traction boundary condition.

## *3.2. Numerical Implementation*

A PINN-based computational solid mechanics program for solving this 1D problem is available at https://github.com/JinshuaiBai/PINN_Comp_Mech. The overall structure of the program is given in Fig. 3. As shown in Fig. 3, the program is divided into three parts, i.e., Pre_Process.py, Train.py, and Post_Porcess.py. The functions in those three files are all called in the Main.py file, as shown in Fig. 4. For the environment settings, PyCharm is selected as the IDE for Python 3.7. Besides, TensorFlow 2.8.0 with Keras 2.8.0 are used to build neural networks. The L-BFGS-B optimiser is provided by SciPy 1.8.0. All the examples are tested on a 64-bit Windows system with an Intel(R) Core(TM) i7-8700 CPU (3.2 GHz).

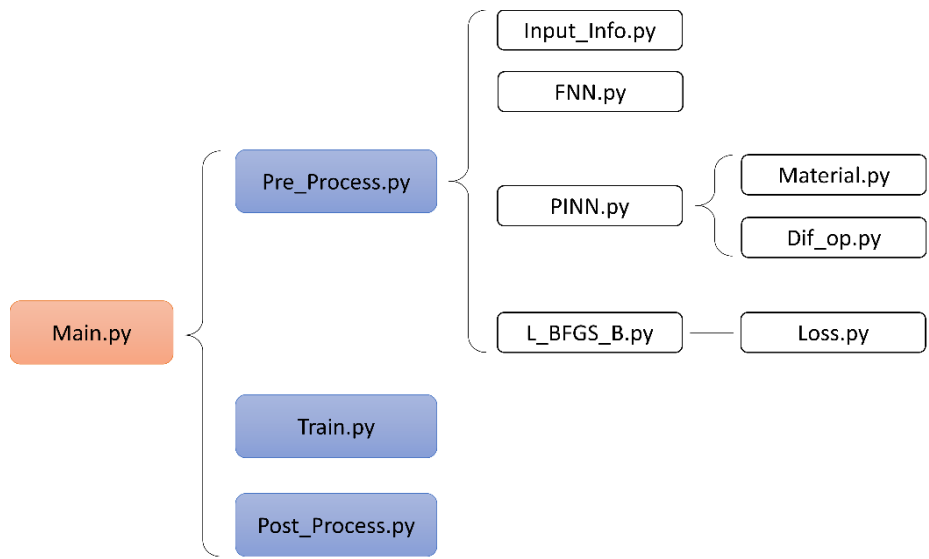

Fig. 3. Program structure for the 1D stretching rod problem.

```
51          """
52              Pre_Process() function is to:
53                  1. Define the geometry of the problem
54                  2. Define the material properties
55                  3. Define the boundary conditions
56                  4. Initialize the neural networks
57                  5. Initialize the optimiser
58          """
59
60          net_u, pinn, l_bfgs_b = Pre_Process()
61
62          """
63              Train() function is to train the PINN with the selected optimiser
64          """
65
66          T, L, it, his_loss = Train(l_bfgs_b)
67
68          """
69              Post_Process() function is to:
70                  1. Visualize the predicted field variables
71                  2. Output results
72          """
73
74          Post_Process(net_u, pinn, his_loss)
```

Fig. 4. The ***Pre_process()**, the **Train()** and the **Post_Process()** functions are called in the Main.py. Those functions are written in the Pre_process.py, the Train.py and the Post_Process.py files, respectively.

3.2.1. Pre-process.py

The ***Pre_process.py*** file defines the problem parameters and initialises the frameworks. Four functions are called in this file, e.g., ***Input_Info()***, ***FNN()***, ***PINN()***, and ***L_BFGS_B()***. In the pre-processing, the problem parameters are firstly loaded through ***Input_Info()*** function. Then, the FNN is built through ***FNN()*** function. With the built FNNs and the loaded problem parameters, the PINN is then constructed through ***PINN()*** function. Finally, the L-BFGS-B optimizer is initialised through the function ***L_BFGS_B()***.

- ***Input_Info():*** This function is to define the problem parameters, including geometrical information, material properties, and neural network settings. In this problem, 51 sample points are generated on the rod with the same spacing. Besides, the FNN parameters in terms of the number of layers, neurons per layer and the activation function are also defined in this function. In this problem, an FNN is used to map the displacement field, $u$, with respect to the coordinate, $x$. The FNN contains three hidden layers, where each layer possesses 5 neurons. The activation function used in the FNN is the tanh function and the initialisation scheme for weights and biases is the LeCun initialisation. The main code of the ***Input_Info()*** function is shown in Fig. 5.

```python
40        ### Define the number of sample points
41        ns = 51
42
43        ### Define the sample points' interval
44        dx = 1./(ns-1)
45
46        ### Initialize sample points' coordinates
47        xy = np.zeros((ns, 1)).astype(np.float32)
48        for i in range(0, ns):
49            xy[i, 0] = i * dx
50        xy_r = np.array([1.])
51
52        ### Create the PINN input list
53        x_train = [xy, xy_r]
54
55        ### Define the Young's modulus
56        E = 10.
57
58        ### Define the traction boundary condition at the right tip of the rod
59        s_r_x = 1.
60
61        ### Create the PINN boundary condition list
62        y_train = [s_r_x]
63
64        ### Define the FNN settings
65        n_input = 1
66        n_output = 1
67        layer = [np.array([5, 5, 5])]
68        acti_fun = 'tanh'
69        k_init = 'LecunNormal'
70        NN_info = [n_input, n_output, layer, acti_fun, k_init]
```

Fig. 5. The *Input_Info()* function defines all the problem parameters and the neural network parameters.

- *FNN()*: This function is to build up an FNN based on the parameters defined in the *Input_Info()* function. We note that the default activation function used for the generated FNN is the tanh function and the default initialisation method for weights and biases is LeCun initialisation. For more options regarding the activation function and initialisation scheme, readers are referred to https://keras.io/api/layers/activations/. The main code of the *FNN()* function is shown in Fig. 6.

```python
31          ### Setup the input layer of the FNN
32          x = tf.keras.layers.Input(shape=(n_input))
33
34          ### Setup the hidden layers of the FNN
35          temp = x
36          for l in layers:
37              temp = tf.keras.layers.Dense(l, activation = acti_fun, kernel_initializer=k_init)(temp)
38
39          ### Setup the output layers of the FNN
40          y = tf.keras.layers.Dense(n_output, kernel_initializer=k_init)(temp)
41
42          ### Combine the input, hidden, and output layers to build up a FNN
43          net = tf.keras.models.Model(inputs=x, outputs=y)
```

Fig. 6. The *FNN()* function builds up an FNN based on the parameters defined in the *Input_Info()* function.

- **PINN()**: This function is to create a PINN with the previously loaded problem parameters and the created FNN. In this function, *Dif()* is first used to obtain the partial differential terms in governing equations. Then, the partial differential terms are fed in *Material()* to calculate strain, stress, and residual from the equilibrium equation. Finally, the stress and residual from the equilibrium equation are output and will be used to formulate the physics-informed loss function. The main code of the *PINN()* function is shown in Fig. 7.

```python
34          ### declare PINN's inputs
35          xy = tf.keras.layers.Input(shape=(1,))
36          xy_r = tf.keras.layers.Input(shape=(1,))
37
38          ### initialize the differential operators
39          Dif_u = Dif(net_u)
40
41          ### obtain the displacment at the right tip of the rod
42          u_r = net_u(xy_r) * xy_r
43
44          ### obtain partial derivatives of u with respect to x
45          u_x, u_xx = Dif_u(xy)
46          u_r_x, u_r_xx = Dif_u(xy_r)
47
48          ### obtain the residuals from the governing equation and traction boundary condition
49          epsilon, sigma, Ge = Material(u_x, u_xx, E)
50          _, sigma_r, _ = Material(u_r_x, u_r_xx, E)
51
52          ### build up the PINN
53          pinn = tf.keras.models.Model(inputs=[xy, xy_r], outputs=[Ge, sigma_r, sigma, epsilon, u_r])
```

Fig. 7. The *PINN()* function create a PINN with the previously loaded problem parameters and the created FNN.

- **L_BFGS_B()**: This function is to initialise the optimiser for training the created FNN. Here, we adopt the L-BFGS-B optimiser, which is a quasi-Newton optimisation algorithm [49]. The physics-informed loss function is employed as the target function for the initialised L-BFGS-B optimiser. Note that the collocation and energy-based loss functions are all prepared in the Loss.py file. To use them, users can load and call the corresponding function in the L-BFGS-B optimiser, as shown in Fig. 9.

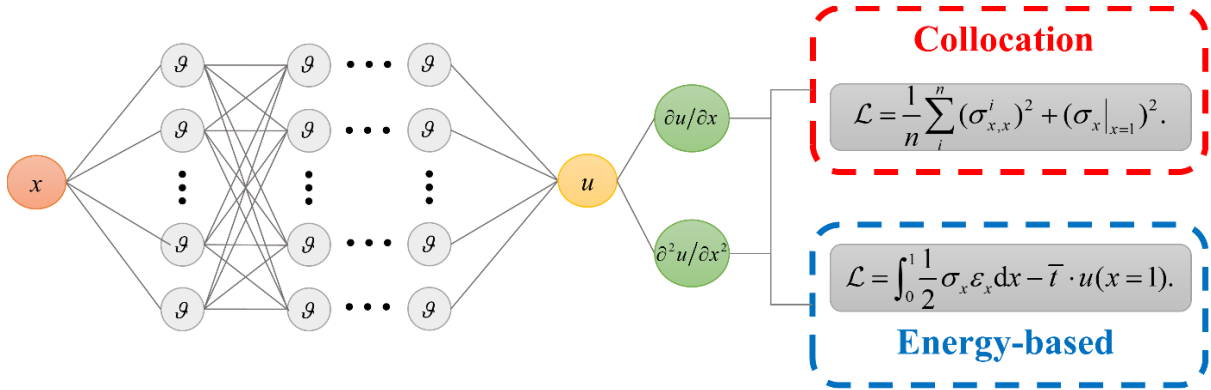

Fig. 8. Physics-informed neural network used for 1D stretching rod problem.

```
138         ### Apply the collocation loss function
139         loss, l1, l2 = Collocation_Loss(y_p, y_train)
140
141         ### Apply the energy-based loss function
142         # loss, l1, l2 = Energy_Loss(y_p, y_train, self.dx)
143
```

Fig. 9. Example of applying different loss functions. As shown in this figure, the program is currently applying the collocation loss function. To implement the energy-based loss function, readers can simply remove line 139 and apply line 142.

3.2.2. Train.py

The **Train.py** file is to execute the training process. The previously prepared optimiser will iteratively minimise the physics-informed loss function till convergence. During the training, the current physics-informed loss will be printed to the command window every 10 training steps (default). Finally, the history of the loss, the final loss, and the number of iterations will be returned after convergence.

3.2.3. Post_process.py

The **Post_process.py** file is to visualise and output the results from the well-trained PINN. Currently, the program provides displacement, strain, and stress plots. Besides, the data in terms of displacement, strain, and stress is also saved in *out.mat* file.

Finally, the flowchart of the program for the 1D stretching rod problem is also given in Table 1.

Table 1. Flowchart of the PINN-based computational solid mechanics program for 1D stretching rod problem.

| **PINN-based computational solid mechanics program for 1D stretching rod** |
|---|
| I. Pre-processing:<br>    i. Load problem information via ***Input_Info()***, including:<br>        Problem geometry;<br>        Material properties;<br>        Boundary conditions;<br>        FNN information.<br>    ii. Build up FNNs via ***FNN()***;<br>    iii. Build up PINN via ***PINN()***;<br>    iv. Initialise the L-BFGS-B optimiser via ***L_BFGS_B()***.<br>II. Training;<br>III. Post-processing. |

## 3.3. Results and discussions

To quantify the error predicted by PINN, we define the relative mean square (RMS) error as follow

$$e_{\text{RMS}} = \frac{1}{n}\sum_{i}^{n}\left(\frac{\omega_i^* - \omega_i^{\text{pred}}}{\omega_{\text{max}}^*}\right)^2. \tag{18}$$

where $\omega^*$ and $\omega^{\text{pred}}$ are ground truth and PINN predicted field variables, respectively. Fig. 10 shows the comparisons of the displacement, strain, and stress results obtained from PINNs with different loss functions and the analytical results. Fig. 11 shows the point-wise absolute error plots from PINNs using different loss functions. It is clear to find that the PINN predicted displacement by using the collocation and energy-based loss function agrees well with the analytical solution. However, the strain and stress predictions by using the energy-based loss function show a large departure from analytical solutions. This is because the energy-based loss function indirectly embeds the equilibrium equation [19]. Consequently, the stress equilibrium equations are not strictly enforced during the training process. Table 2 compares the RMS errors, the number of iterations required for convergence and the CPU time while utilising different loss functions. The PINN trained with the collocation loss function achieves higher accuracy but converges more slowly than the PINN trained with the energy-based loss function. Compared to the collocation loss function, which contains second-order differential terms, the energy-based loss function only contains first-order differential terms. The use of lower-order differential physics makes the energy-based loss function easier to numerically implement. Therefore, PINNs with the energy-based loss function are computationally more efficient than PINNs with the collocation loss function.

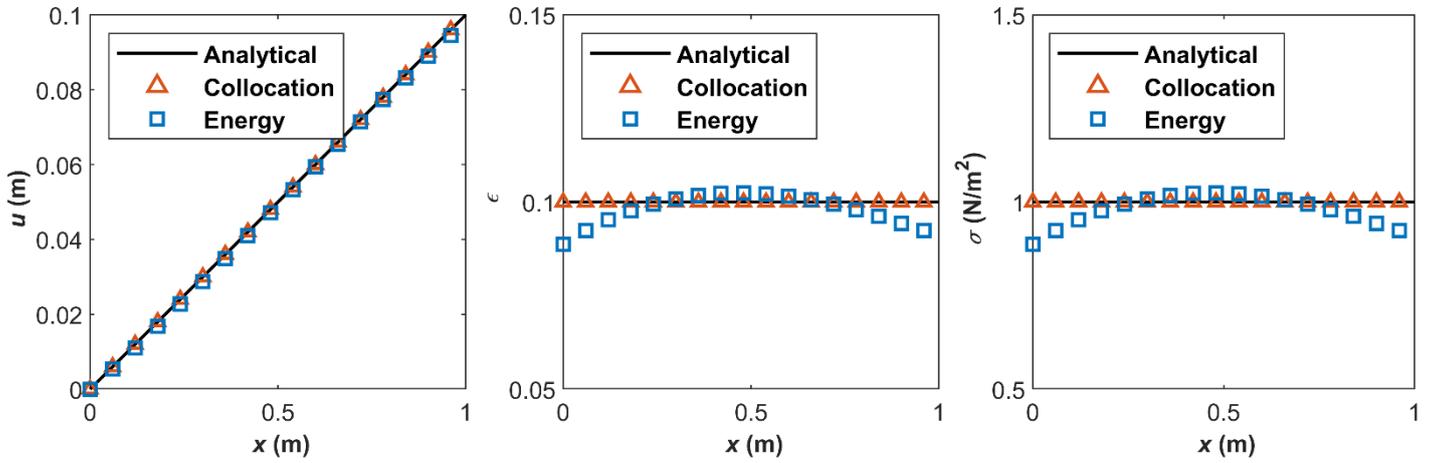

Fig. 10. Comparisons of the displacement, strain, and stress fields between the collocation loss function, energy-based loss function and the analytical solution.

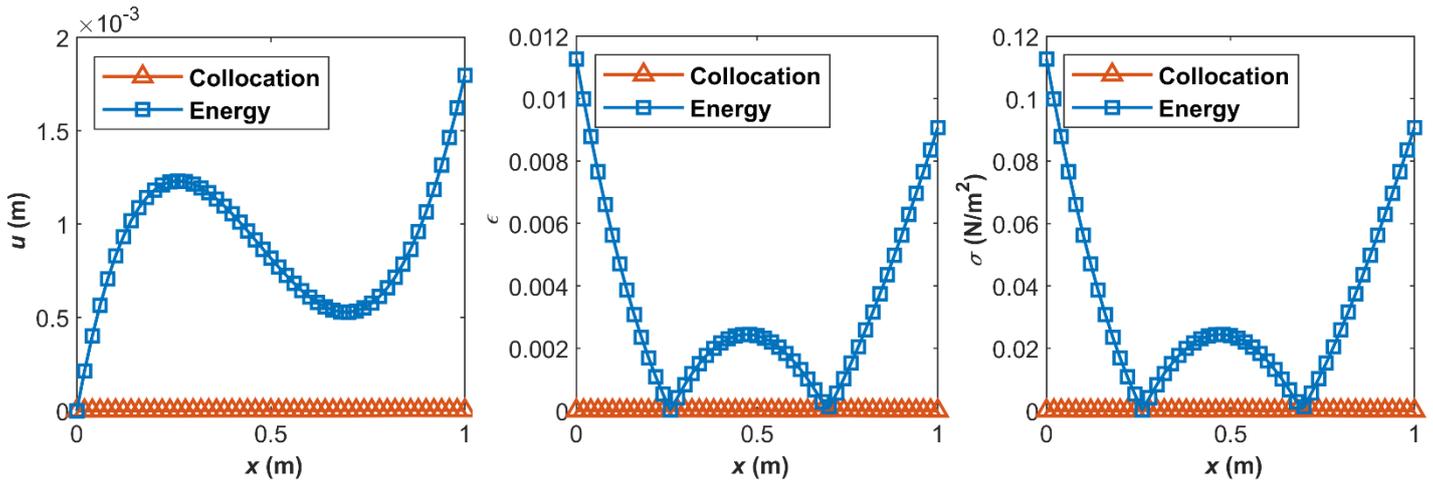

Fig. 11. Point-wise absolute error plots of the displacement, strain, and stress fields from PINNs using the collocation loss function and the energy-based loss function.

Table 2. Comparisons of the RMS errors, number of iterations for convergence and the CPU time with respect to different kinds of loss functions for the 1D problem.

|  | RMS errors | | | Iterations | CPU time (s) |
| --- | --- | --- | --- | --- | --- |
|  | $U$ | $\varepsilon$ | $\sigma$ | | |
| Collocation | $5.20 \times 10^{-10}$ | $1.96 \times 10^{-9}$ | $1.96 \times 10^{-9}$ | 83 | 1.0996 |
| Energy-based | $1.80 \times 10^{-3}$ | $1.70 \times 10^{-3}$ | $1.70 \times 10^{-3}$ | 39 | 0.6559 |

## 4. PINN-based Computational Solid Mechanics in Two Dimension

*4.1. In-plain stretching square plate*

In this section, the program is extended to a 2D in-plain stretching square plate problem. The configuration of the problem is shown in Fig. 12(a). Due to the geometric symmetry, a quarter of the plate is used for the modelling, which is shown in Fig. 12(b). The length of the plate $L = 2$ m. A distributed force, $\bar{t}(y)$, is applied on the right side of the plate

$$\bar{t}(y) = \cos(\frac{\pi y}{2}). \tag{19}$$

The displacement boundary conditions are given as follows

$$U(0, y) = 0, V(x, 0) = 0. \tag{20}$$

For the plane stress problem, the Lamé constants are obtained through

$$\begin{cases} \lambda = \dfrac{E\upsilon}{(1+\nu)(1-\nu)}, \\ \mu = \dfrac{E}{2(1+\nu)}, \end{cases} \tag{21}$$

where $E = 7$ Pa and $v = 0.3$ are used in the plate problems.

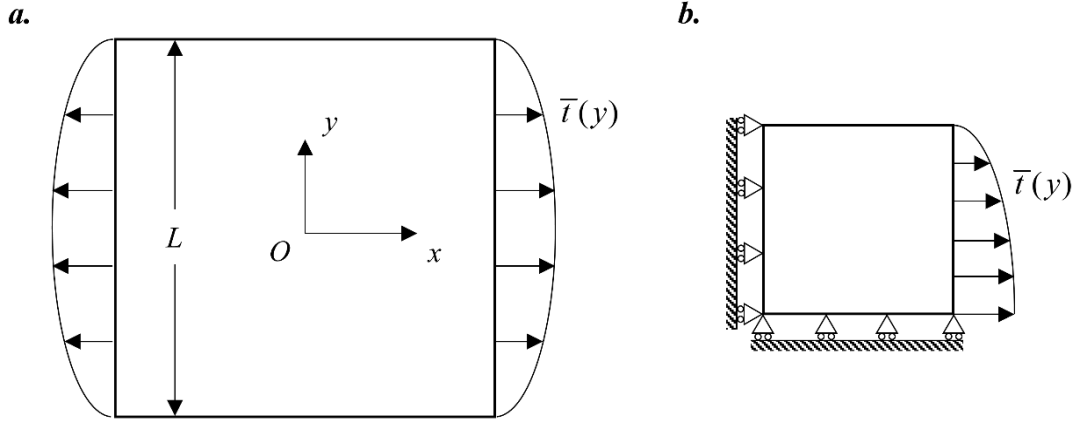

Fig. 12. (a) Configuration of the in-plain stretching square plate problem; (b) Actual geometry applied for modelling.

*4.2. Numerical implementation*

In the computational domain, 2601 uniformly distributed sample points are generated in the computational domain, where the spacing of sample points in all directions is 0.02 m. Furthermore, two FNNs with the same structures are applied to respectively predict $U$ and $V$, where each FNN contains three hidden layers and 20 neurons per hidden layer, as shown in Fig. 13. The FEM results from ABAQUS with fine mesh are used as the reference, where the quadratic element is applied, and the size of all elements is 0.005 m.

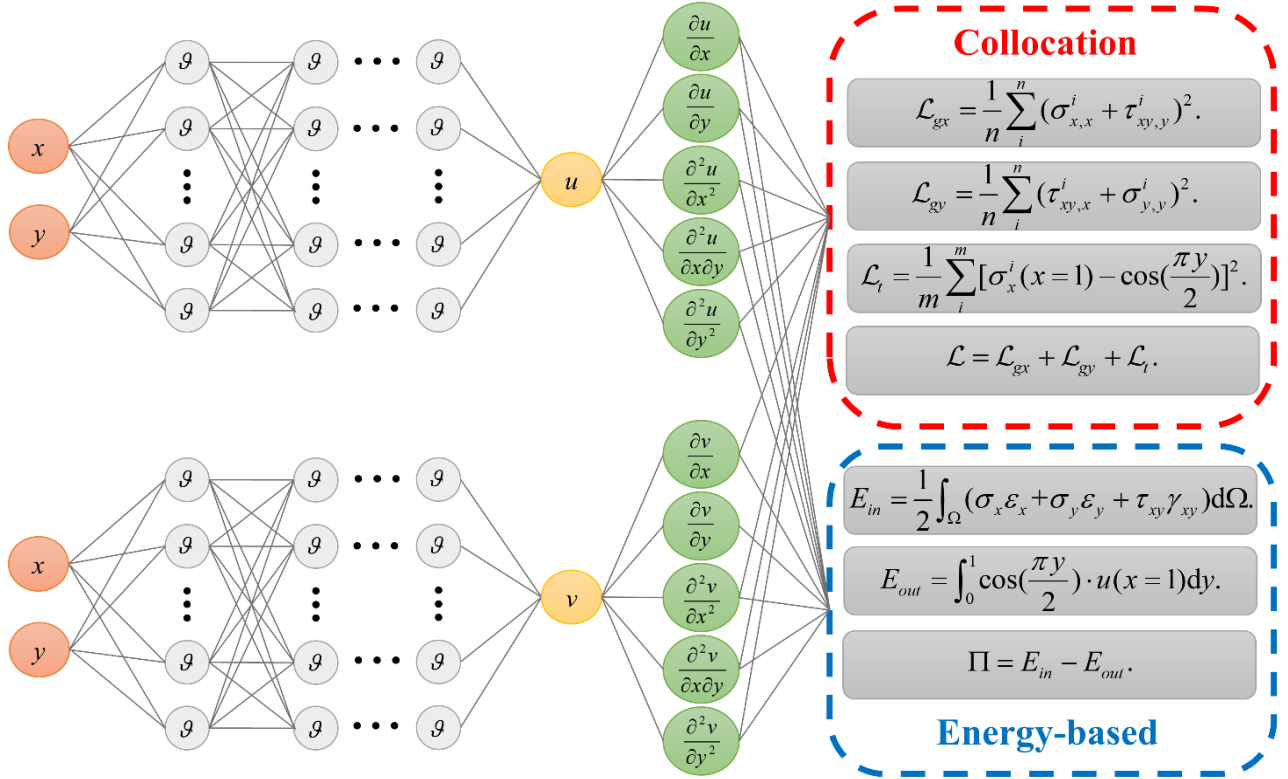

Fig. 13. Physics-informed neural networks used for 2D in-plain stretching plate. Two FNNs are applied to predict *U* and *V*, respectively.

### 4.3. Results and discussions

Table 3 shows the comparisons of the RMS errors, the number of iterations for convergence and the CPU time by using different loss functions. As observed from the table, the PINN with collocation loss function achieves higher accuracy than the PINN with energy-based loss function. However, the PINN with the energy-based loss function converges significantly earlier than that with the collocation loss function. As for computational efficiency, the training time for the PINN with the energy-based loss function is significantly faster than the training time for the PINN with the collocation loss function. However, the computational efficiency of PINN is still incomparable with the prevailing used computational mechanics methods. This is because PINNs are complex nonlinear systems and can be only trained by gradient descendent algorithms, while the prevailingly used computational mechanics methods can simplify the governing equations into a linear system (i.e. KU = F). Fig. 14 shows the PINNs' predictions, and FEM reference results and Fig. 15 shows the point-wise error contours of the displacement and stress fields for the in-plane stretching problem. The FEM results are given in Fig. 14(c) for reference. As observed from Fig. 14(a), the PINN with the collocation loss function successfully predicts all the field variables well, while only a minor discrepancy is found. Besides, using the collocation loss function, PINN clearly captures the stress concentration of $\sigma_y$ at the bottom right corner of the plate. As for the energy-based loss function results shown in Fig. 14(b), the displacement fields show good agreement with the FEM results. However, obvious oscillations can be observed in the three stress fields contour obtained by the energy-based loss function. This is mainly because the energy-based loss function lacks direct information on the equilibrium equation. Since the stress fields are calculated through the derivatives of the displacement fields, the stress fields can be sensitive even for slight displacement errors. Nevertheless, the stress fields from the energy-based loss function roughly capture the patterns of the stresses from the FEM. In this manner, post-processing methods can be applied to further smoothen the predictions from the energy-based loss function. The above conclusions are further demonstrated by the error contours. Fig. 16 shows the comparisons of the displacement and stress fields along the diagonal of the plate from position (0,1) to (1,0). It is clear that the displacement predictions from both collocation and energy-based loss functions perform well, as illustrated in Fig. 16(a). For the stress fields shown in Fig. 16(b), the collocation results align with the reference results, while the $\sigma_x$ and $\sigma_y$ from the energy-based loss function show discrepancy near the boundary of the plate. More investigations and efforts are required to improve the accuracy of the energy-based loss function for stress fields, especially the enforcement of boundary conditions.

Table 3. Comparisons of the RMS errors, the number of iterations for convergence and the CPU time with respect to different kinds of loss functions for the 2D problem.

|  | RMS errors | | | | | Iterations | CPU time (s) |
| --- | --- | --- | --- | --- | --- | --- | --- |
|  | $U$ | $V$ | $\sigma_x$ | $\sigma_y$ | $\tau_{xy}$ |  |  |
| Collocation | $3.24 \times 10^{-7}$ | $3.06 \times 10^{-6}$ | $6.28 \times 10^{-6}$ | $2.84 \times 10^{-5}$ | $1.86 \times 10^{-4}$ | 2491 | 62.4290 |
| Energy | $2.61 \times 10^{-4}$ | $2.68 \times 10^{-3}$ | $2.16 \times 10^{-3}$ | $4.03 \times 10^{-3}$ | $9.14 \times 10^{-2}$ | 721 | 6.2549 |

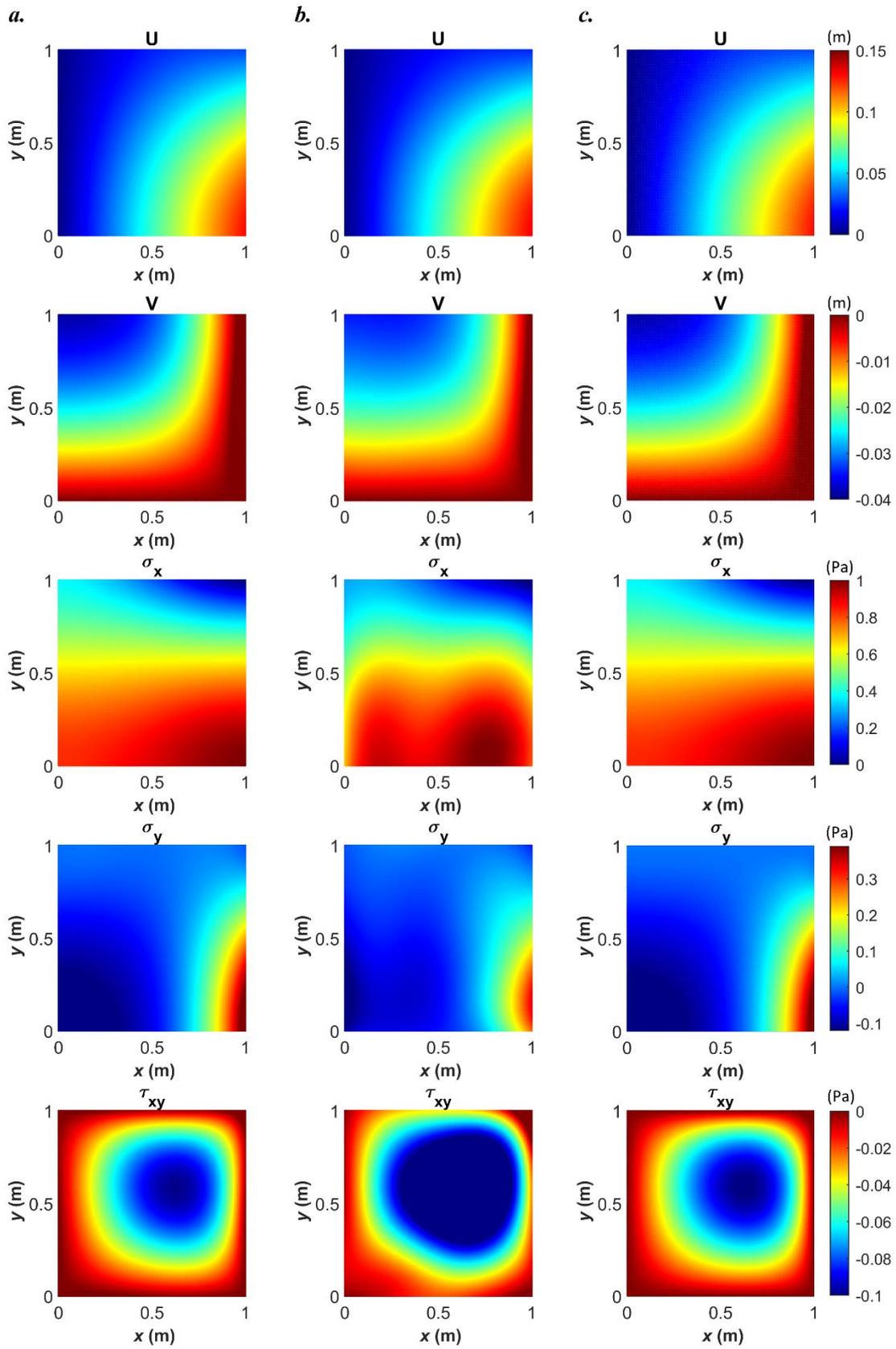

Fig. 14. Comparison of the displacement and stress fields of the in-plain stretching square plate problem from (a) collocation loss function; (b) Energy-based loss function; (c) FEM.

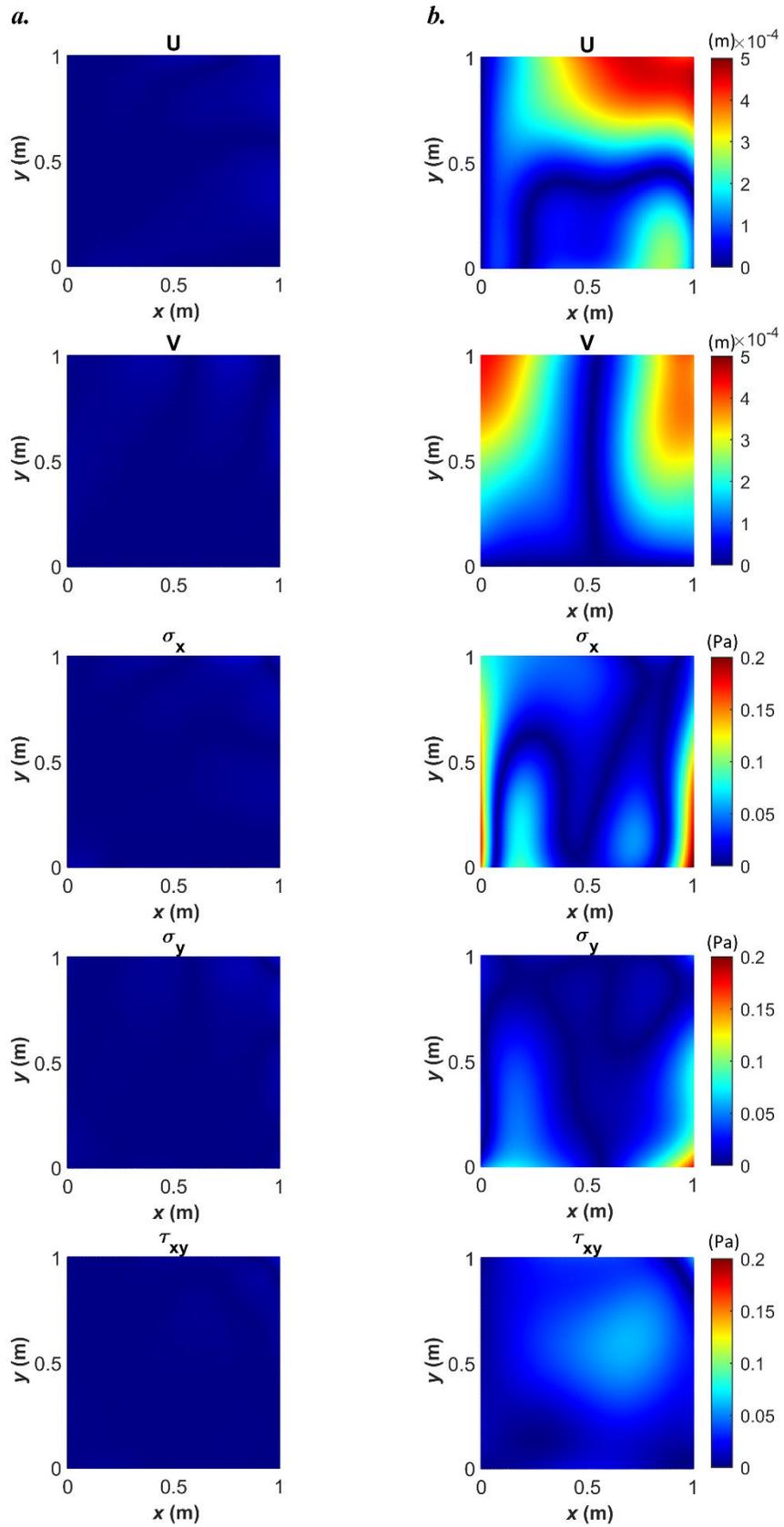

Fig. 15. (a) Point-wise absolute error contours from PINN using the collocation loss function. (b) Point-wise absolute error contours from PINN using the collocation loss function.

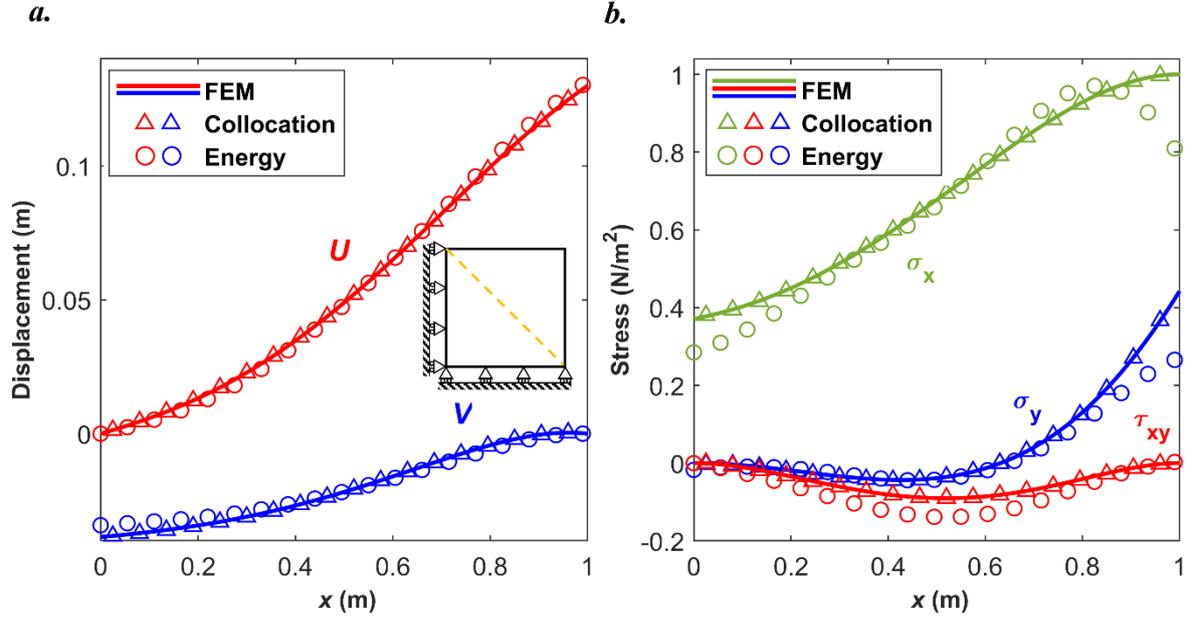

Fig. 16. (a) Comparison of the displacement along the diagonal of the plate, from (0,1) to (1,0); (b) Comparison of the stress fields along the diagonal of the plate, from (0,1) to (1,0). The solid lines in the two figures denote the results from the FEM. The triangles and circles in the two figures respectively denote the results from the collocation loss function and energy-based loss function.

## 5. PINN-based Computational Solid Mechanics in Three Dimension

### 5.1. Stretching cube

A three-dimensional stretching cube problem is conducted here. The configuration of the problem is shown in Fig. 17(a). The length of the cube $L = 2$ m. A distributed force, $\bar{t}(x,y)$, is applied on the top and bottom surfaces of the cube in the $z$ direction

$$\bar{t}(x,y) = \cos(\frac{\pi x}{2})\cos(\frac{\pi y}{2}). \tag{22}$$

The displacement boundary conditions are given as follows

$$U(0,y,z) = 0, V(x,0,z) = 0, W(x,y,0) = 0. \tag{23}$$

The Young's modulus and Poisson's ratio are $E = 10$ Pa and $v = 0.25$. Due to its symmetrical property, one-eighth of the geometry is modelled, which is shown in Fig. 17(b).

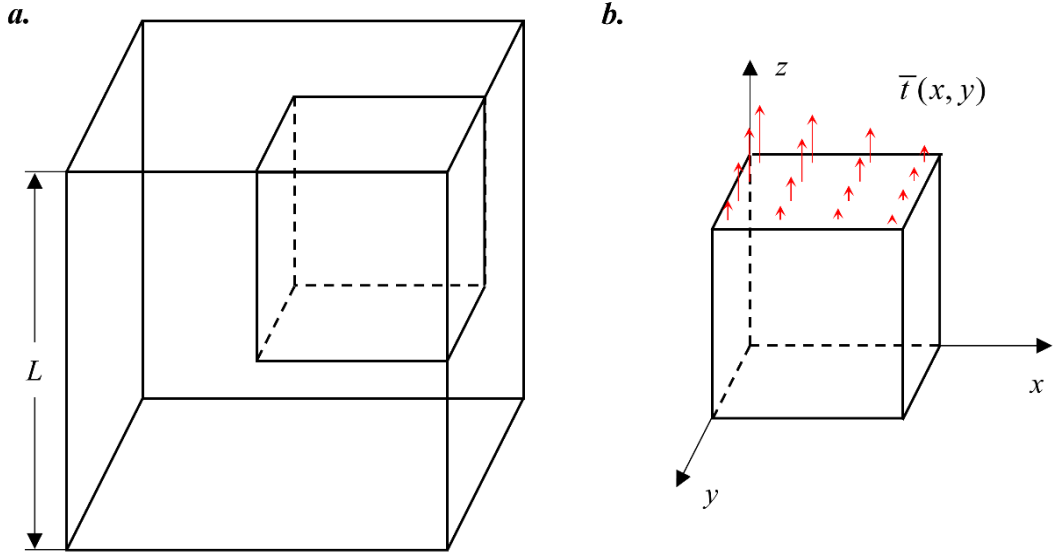

Fig. 17. (a) Configuration of the stretching rod cube problem; (b) Actual geometry used in modelling due to the geometric symmetry. The arrows in red denote the distributed force boundary condition.

*5.2. Numerical implementation*

Overall 9261 sample points are generated in the computational domain, where the spacing of sample points in all directions is 0.05m. Three FNNs are established for predicting *U*, *V* and *W*, respectively. Each FNN contains 4 layers and 20 neurons per hidden layer. Since the analytical solution is not available for this problem, the FEM results from ABAQUS with a fine mesh are used as the reference, where the quadratic element is applied and the size of all elements is 0.01 m. Here, only the collocation loss function is applied to deal with this problem. Readers can easily implement the energy-based loss function through the collocation program with small modifications.

*5.3. Results and discussions*

Fig. 18 and Fig. 19 show the comparisons of the displacement and stress fields of the problem between the proposed loss function and FEM results, respectively. From the figures, one can find that both the displacement and stress fields agree with the reference results. Besides, results from PINN-based computational solid mechanics show good symmetrical properties. However, Fig. 20 shows the point-wise absolute error contours from PINNs by using the collocation loss function. It is clear to observe that the error for displacement *U* is relatively larger than the error for displacement *V*. This is because three independent FNNs are applied to respectively predict the displacement fields and are trained separately. Given that the displacement *U* and *V* should be symmetry to each other. Thus, one FNN can be applied for predicting both the displacement *U* and *V*, instead of using two independent FNNs. Besides, large point-wise absolute errors can be always observed at the corner of the cube, suggesting that more attention should be paid to minimising the residuals calculated from corners. We note that adaptive learning strategies [15, 50] can be a way to address this issue.

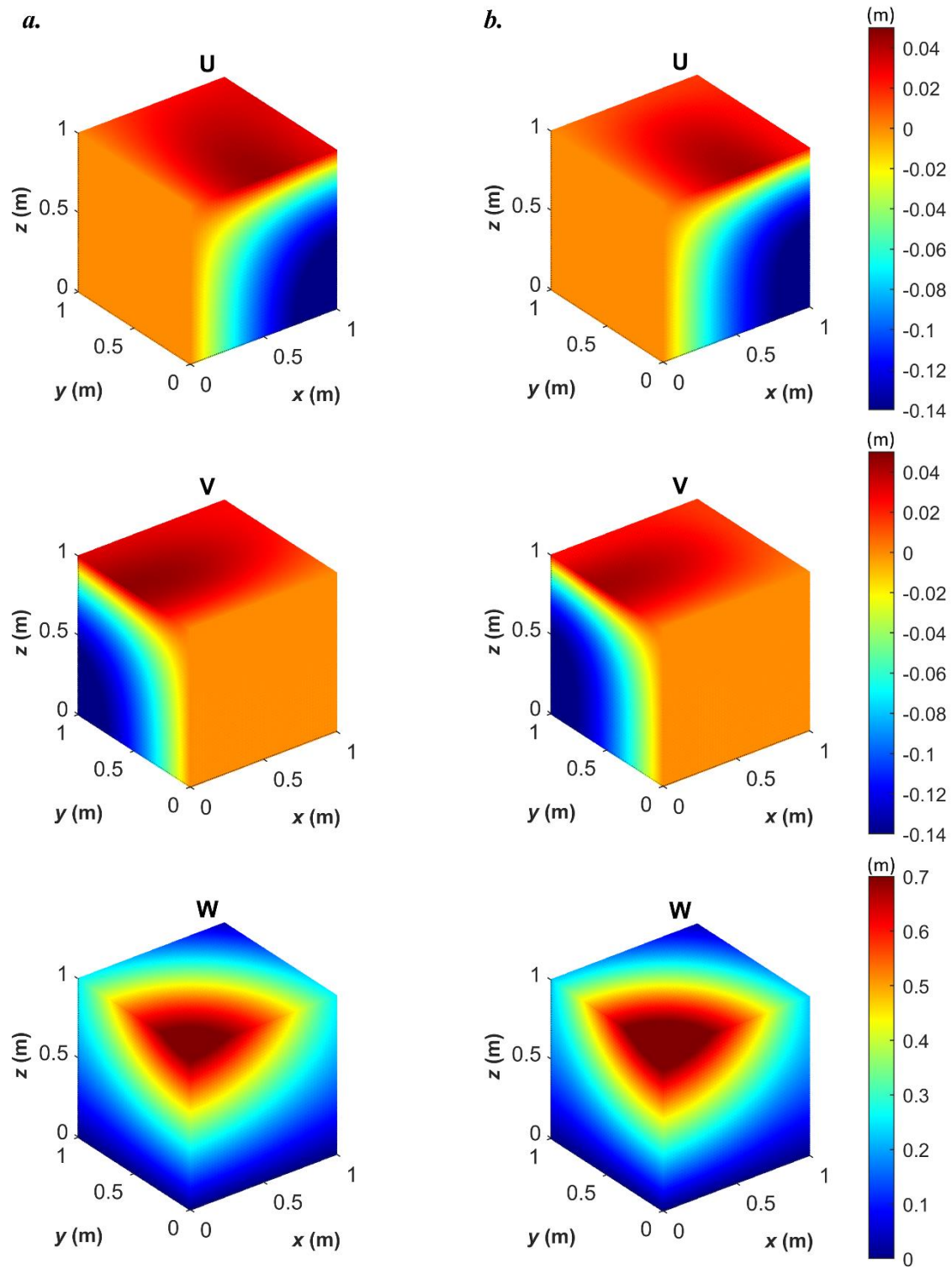

Fig. 18. Comparisons of the displacement fields of the 3D stretching cube problem from (a) PINN-based computational solid mechanics; (b) FEM.

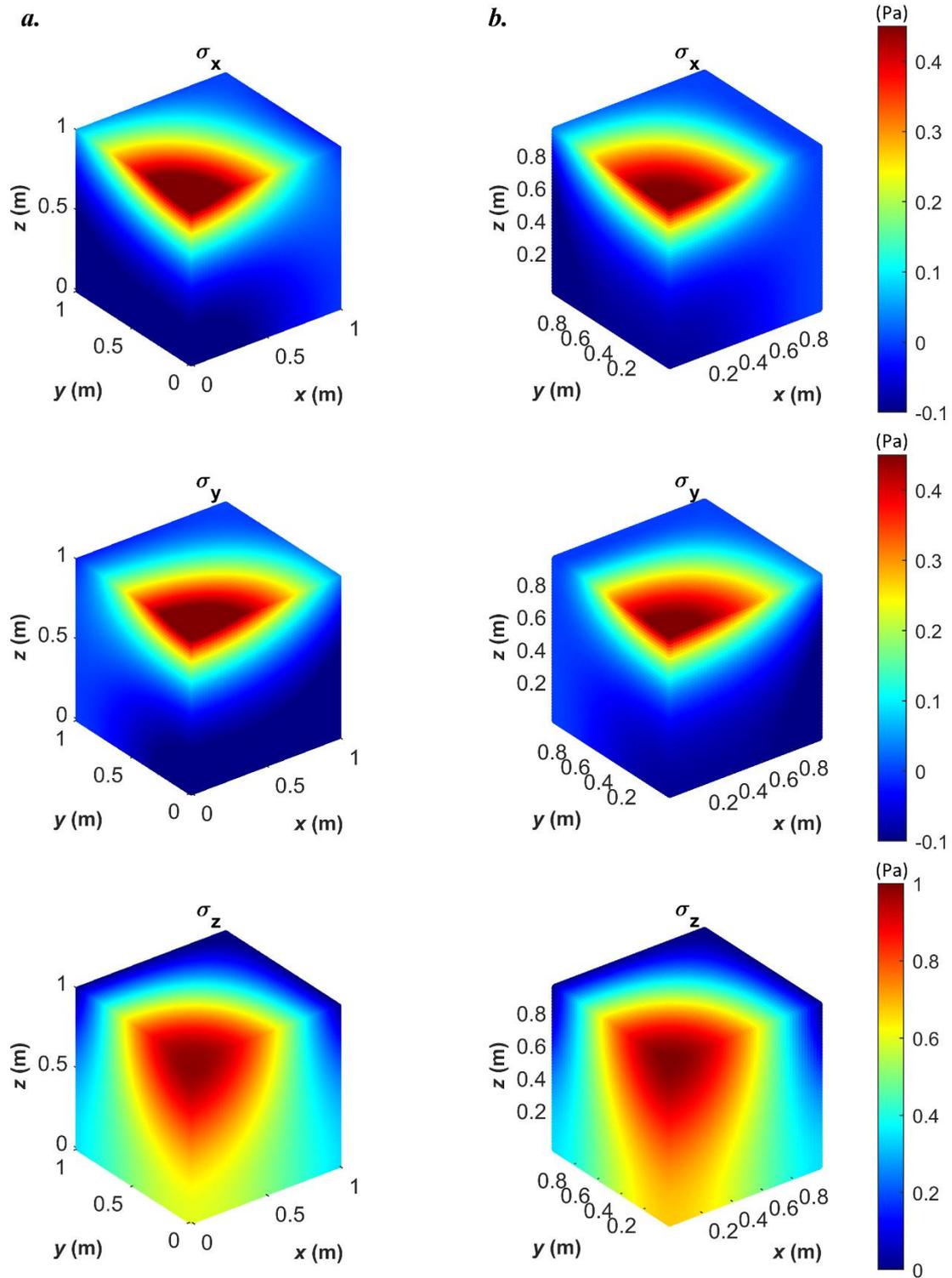

Fig. 19. Comparisons of the stress fields of the 3D stretching cube problem from (a) PINN-based computational solid mechanics; (b) FEM.

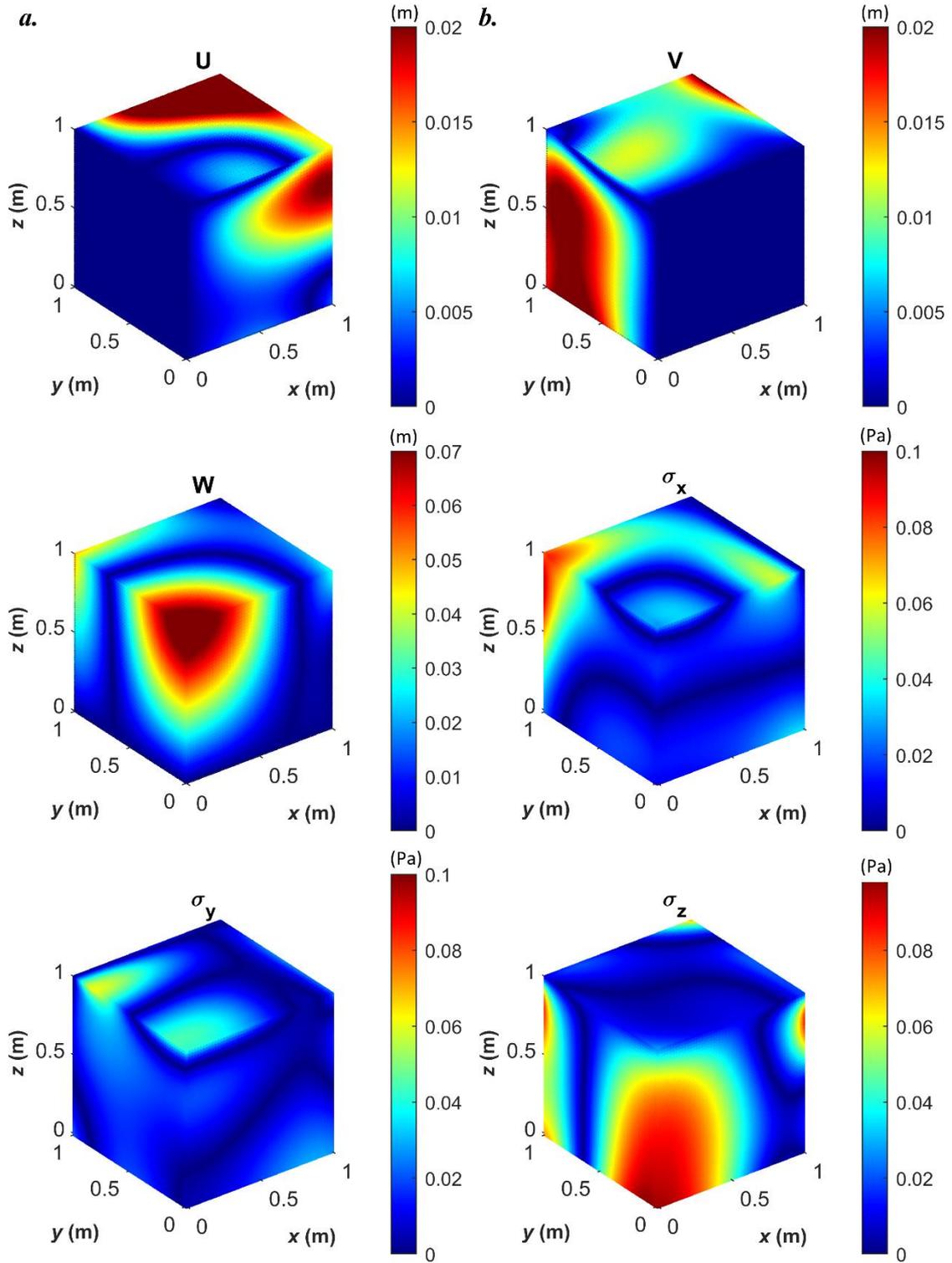

Fig. 20. Point-wise absolute error contours from the PINN using the collocation loss function.

## 6. Summary

In this paper, details of the PINN-based computational solid mechanics and its numerical implementation have been introduced. The prevailingly used physics-informed loss functions for PINN-based computational solid mechanics are summarised. Additionally, examples including 1D, 2D, and 3D problems are presented to show the performance and capability of PINN-based computational solid mechanics. It has been demonstrated that both types of physics-informed loss functions are effective for displacement predictions. The PINN with the collocation loss function can achieve better accuracy for stress fields since the equilibrium equation is explicitly imposed in the loss function. In contrast, PINN with the energy-based loss function is computationally more efficient, for it requires lower-order differential terms and is easier to implement. However, the use of the energy-based loss function can result in severe stress prediction errors. This is because the energy-based loss function does not enforce the equilibrium equation explicitly. Furthermore, programs based on the Python coding language are provided with step-by-step explanations. It is worth noting that the programs for the PINN-based computational solid mechanics are manoeuvrable and can be easily extended to broaden applications, such as geometric nonlinearity and hyperelastic problems. Additionally, with the programs, more investigations regarding the neural network settings, training algorithms sections and sample points initialisation strategies can be easily implemented. We remain this to the readers to explore.

Despite its good performance, PINN-based computational mechanics is still in its infancy. Issues regarding robustness and computational efficiency are still severe during the use of PINN-based computational mechanics [20]. Besides, the number of hidden layers and neurons is mainly determined by the authors' experiences. Currently, only empirical criteria are available to determine the size of neural networks for various applications [41, 51]. More investigations should be conducted to determine the size of neural networks from the computational mechanics regard. This work aims to provide readers with a fundamental insight into PINN-based computational mechanics. Meanwhile, we provide open questions regarding the effectiveness, robustness and efficiency of PINN-based computational solid mechanics. The great potential of PINN-based computational mechanics still remains to be tapped. We hope this work can spark further investigations and development of PINN-based computational mechanics to be an effective way for mechanics applications.

## Author Contributions

**J. Bai:** Conceptualization, Methodology, Coding, Formal analysis, Writing-Original draft. **H. Jeong:** Coding, Writing-Reviewing and Editing. **C. P. Batuwatta-Gamage:** Coding, Writing-Reviewing and Editing. **S. Xiao:** Coding, Writing-Reviewing and Editing. **Q. Wang:** Writing-Reviewing and Editing. **C.M. Rathnayaka:** Writing-Reviewing and Editing. **L. Alzubaidi:** Writing-Reviewing and Editing. **G. Liu:** Writing-Reviewing and Editing. **Y. Gu:** Conceptualization, Methodology, Formal analysis, Writing-Reviewing and Editing, Supervision.

## Declaration of Competing Interest

The authors declare no competing interests.

## Acknowledgements

Support from the Australian Research Council research grants (IC190100020 and DP200102546) is gratefully acknowledged (J. Bai).

## Reference

1.  Liu G-R (2009) Meshfree methods: moving beyond the finite element method CRC press

2.  Liu G-R and Quek SS (2013) The finite element method: a practical course Butterworth-Heinemann

3.  Zhang Y, Gao Z, Wang X and Liu Q (2022) Predicting the pore-pressure and temperature of fire-loaded concrete by a hybrid neural network. International Journal of Computational Methods 19: 2142011.

4.  Zhang Y, Gao Z, Wang X and Liu Q (2023) Image Representations of Numerical Simulations for Training Neural Networks. Computer Modeling in Engineering & Sciences 10.32604/cmes.2022.022088

5.  Khan MIH, Batuwatta-Gamage CP, Karim M and Gu Y (2022) Fundamental Understanding of Heat and Mass Transfer Processes for Physics-Informed Machine Learning-Based Drying Modelling. Energies 15: 9347.

6.  Li Z, Bai J, Ouyang H, Martelli S, Tang M, Wei H, Liu P, Han W-R and Gu Y (2023) Physics-informed neutral network for friction-involved nonsmooth dynamics problems. arXiv preprint arXiv:02542

7.  Bai J, Zhou Y, Rathnayaka CM, Zhan H, Sauret E and Gu Y (2021) A data-driven smoothed particle hydrodynamics method for fluids. Engineering Analysis with Boundary Elements 132: 12-32.

8.  Zhou Y, Zhan HF, Zhang WH, Zhu JH, Bai JS, Wang QX and Gu YT (2020) A new data-driven topology optimization framework for structural optimization.

9.  Raissi M, Yazdani A and Karniadakis GE (2020) Hidden fluid mechanics: Learning velocity and pressure fields from flow visualizations. Science 367: 1026-1030.

10. Liu G-R (2022) Machine Learning with Python: Theory and Applications World Scientific

11.	Raissi M, Perdikaris P and Karniadakis GE (2019) Physics-informed neural networks: A deep learning framework for solving forward and inverse problems involving nonlinear partial differential equations. Journal of Computational Physics 378: 686-707.

12.	Meng X, Li Z, Zhang D and Karniadakis GE (2020) PPINN: Parareal physics-informed neural network for time-dependent PDEs. Computer Methods in Applied Mechanics and Engineering 370 10.1016/j.cma.2020.113250

13.	Dong S and Li Z (2021) Local extreme learning machines and domain decomposition for solving linear and nonlinear partial differential equations.

14.	Wang S, Teng Y and Perdikaris P (2020) Understanding and mitigating gradient pathologies in physics-informed neural networks. arXiv preprint arXiv:04536

15.	Wang S, Yu X and Perdikaris P (2021) When and why PINNs fail to train: A neural tangent kernel perspective.

16.	Sukumar N and Srivastava A (2021) Exact imposition of boundary conditions with distance functions in physics-informed deep neural networks.

17.	Yang L, Meng X and Karniadakis GE (2021) B-PINNs: Bayesian physics-informed neural networks for forward and inverse PDE problems with noisy data.

18.	Kharazmi E, Zhang Z and Karniadakis GEM (2021) hp-VPINNs: Variational physics-informed neural networks with domain decomposition.

19.	Fuhg JN and Bouklas N (2021) The mixed Deep Energy Method for resolving concentration features in finite strain hyperelasticity.

20.	Karniadakis GE, Kevrekidis IG, Lu L, Perdikaris P, Wang S and Yang L (2021) Physics-informed machine learning. Nature Reviews Physics 3: 422-440.

21.	Haghighat E, Raissi M, Moure A, Gomez H and Juanes R (2021) A physics-informed deep learning framework for inversion and surrogate modeling in solid mechanics.

22.	Samaniego E, Anitescu C, Goswami S, Nguyen-Thanh VM, Guo H, Hamdia K, Zhuang X and Rabczuk T (2020) An energy approach to the solution of partial differential equations in computational mechanics via machine learning: Concepts, implementation and applications.

23.	Goswami S, Yin M, Yu Y and Karniadakis GE (2022) A physics-informed variational DeepONet for predicting crack path in quasi-brittle materials.

24.	Zhuang X, Guo H, Alajlan N, Zhu H and Rabczuk T (2021) Deep autoencoder based energy method for the bending, vibration, and buckling analysis of Kirchhoff plates with transfer learning.

25.	Nguyen-Thanh VM, Anitescu C, Alajlan N, Rabczuk T and Zhuang X (2021) Parametric deep energy approach for elasticity accounting for strain gradient effects.

26.	Li W, Bazant MZ and Zhu J (2021) A physics-guided neural network framework for elastic plates: Comparison of governing equations-based and energy-based approaches.

27.	Rao C, Sun H and Liu Y (2021) Physics-Informed Deep Learning for Computational Elastodynamics without Labeled Data.

28.	Haghighat E, Bekar AC, Madenci E and Juanes R (2021) A nonlocal physics-informed deep learning framework using the peridynamic differential operator.

29. Abueidda DW, Lu Q and Koric S (2021) Meshless physics-informed deep learning method for three-dimensional solid mechanics. International Journal for Numerical Methods in Engineering

30. Wessels H, Weißenfels C and Wriggers P (2020) The neural particle method – An updated Lagrangian physics informed neural network for computational fluid dynamics.

31. Cai S, Wang Z, Fuest F, Jeon YJ, Gray C and Karniadakis GE (2021) Flow over an espresso cup: inferring 3-D velocity and pressure fields from tomographic background oriented Schlieren via physics-informed neural networks.

32. Bai J, Zhou Y, Ma Y, Jeong H, Zhan H, Rathnayaka C, Sauret E and Gu Y (2022) A general Neural Particle Method for hydrodynamics modeling.

33. Mojahedin A, Hashemitaheri M, Salavati M, Fu X and Rabczuk T (2022) A Deep Energy Method for the Analysis of Thermoporoelastic Functionally Graded Beams. International Journal of Computational Methods 10.1142/s0219876221430209

34. Zhang E, Dao M, Karniadakis GE and Suresh S (2022) Analyses of internal structures and defects in materials using physics-informed neural networks. Science Advances 8: eabk0644.

35. Abadi M, Agarwal A, Barham P, Brevdo E, Chen Z, Citro C, Corrado GS, Davis A, Dean J and Devin M (2016) Tensorflow: Large-scale machine learning on heterogeneous distributed systems.

36. Bai J, Rabczuk T, Gupta A, Alzubaidi L and Gu Y (2022) A physics-informed neural network technique based on a modified loss function for computational 2D and 3D solid mechanics.

37. Nguyen-Thanh VM, Zhuang X and Rabczuk T (2020) A deep energy method for finite deformation hyperelasticity.

38. Goswami S, Anitescu C, Chakraborty S and Rabczuk T (2020) Transfer learning enhanced physics informed neural network for phase-field modeling of fracture.

39. Liu G-R and Gu Y (2005) An introduction to meshfree methods and their programming Springer Science & Business Media

40. Bengio Y, Goodfellow I and Courville A (2017) Deep learning MIT press Massachusetts, USA:

41. LeCun Y, Bengio Y and Hinton G (2015) Deep learning. Nature 521: 436-444.

42. Hornik K, Stinchcombe M and White H (1989) Multilayer feedforward networks are universal approximators. Neural Networks 2: 359-366.

43. Henkes A, Wessels H and Mahnken R (2022) Physics informed neural networks for continuum micromechanics.

44. Nielsen MA (2015) Neural networks and deep learning Determination press San Francisco, CA

45. Nwankpa C, Ijomah W, Gachagan A and Marshall S (2018) Activation functions: Comparison of trends in practice and research for deep learning.

46. Baydin AG, Pearlmutter BA, Radul AA and Siskind JM (2018) Automatic differentiation in machine learning: a survey. Journal of Machine Learning Research 18

47.     McClenny L and Braga-Neto U (2020) Self-adaptive physics-informed neural networks using a soft attention mechanism.

48.     Lanczos C (2020) The variational principles of mechanics University of Toronto press

49.     Morales JL and Nocedal J (2011) Remark on "Algorithm 778: L-BFGS-B: Fortran subroutines for large-scale bound constrained optimization". ACM Transactions on Mathematical Software 38: 1-4.

50.     Wang S, Wang H and Perdikaris P (2021) On the eigenvector bias of Fourier feature networks: From regression to solving multi-scale PDEs with physics-informed neural networks. Computer Methods in Applied Mechanics and Engineering 384 10.1016/j.cma.2021.113938

51.     Bengio Y, Lamblin P, Popovici D and Larochelle H (2007) Greedy layer-wise training of deep networks  Advances in neural information processing systems, pp 153-160.